\input{aipcheck}

\documentclass[
    ,final            
  ]
  {aipproc}

\layoutstyle{6x9}

\begin{document}

\title{Neutron Spin Structure Measurements in JLab Hall A}

\author{Jian-ping Chen, for the JLab Hall A and E94-010 Collaborations}
{
  address={Jefferson Lab, Newport News, Virginia 23606, USA}
}


\begin{abstract}
Recent progress from Jefferson Lab has significantly improved our 
understanding of the nucleon spin structure in the high-$x$ region. 
Results from two experiments in Hall A are presented. The first 
experiment is a precision measurement of the neutron spin asymmetry,
$A_1^n$, in the high-$x$ (valence quark) region. 
The results show for the first time that $A_1^n$ becomes positive at 
large $x$, strongly breaking SU(6) (spin-flavor) symmetry. The data 
trend is in good agreement with SU(6)-breaking valence quark models 
and with the fits to the previous world data. 
Combining the $A_1^n$ results with the world $A_1^p$ data, the 
up and down quark spins distributions in the nucleon were extracted. 
The results showed
that for the proton the valence down quark spin is in the opposite direction 
from that of the proton, in disagreement with predictions of leading-order 
perturbative QCD models, which neglect quark orbital angular momentum. 

In the second experiment, the $Q^2$ dependence of the moments and 
sum rules of the spin structure in the low to intermediate $Q^2$ region
were measured,  
providing a unique bridge linking the quark-gluon picture of the nucleon 
and the coherent hadronic picture.

\end{abstract}

\maketitle

\section{Introduction}

Since the `spin crisis'~\cite{spin}, substantial efforts, both theoretical and 
experimental, have been devoted to understanding the nucleon's spin puzzle. 
A new generation of experiments were carried out in the 1990s
at SLAC, CERN and DESY. These experiments concluded that the quarks carry 
about $ 20-30\%$ of the nucleon spin. The rest of the nucleon spin should come 
from the quark orbital angular momentum (OAM) and the gluon total angular 
momentum. There is almost no direct experimental information available on the 
quark OAM and the gluon total angular momentum.
The Bjorken sum rule\cite{Bjorken}, a fundamental sum rule of the spin 
structure function based on QCD, was verified to an accuracy of
better than $10\%$. Attempts have been made to extract the parton spin
distributions from global analyses of the polarized 
deep-inelastic-scattering data. The uncertainties are much larger than those 
of the unpolarized parton distribution due to the fact that the polarized 
data coverage is much less extensive than that of the unpolarized data.   

Recently, the high polarized luminosity available at Jefferson 
Lab (JLab) has allowed the study of the nucleon spin structure at 
an unprecedented precision, enabling us to access the hard-to-reach 
valence quark (high-$x$) region and also to accurately map the 
intermediate to low $Q^2$ region. 

The high-$x$ region is of special interest, because this is where the valence 
quark contributions are expected to dominate.
With sea quarks and explicit gluon contributions expected not to be
important, it is a clean region to test our understanding of nucleon
structure. Relativistic constituent quark models~\cite{vqm}
should be applicable in this region
and perturbative QCD~\cite{pQCD} is also able to make predictions in the large
$x$ ($x \rightarrow ~ 1$) limit. JLab experiment 
E99-117 measured, with high precision, the spin asymmetry $A_1^n$ in the 
high-$x$ region and extracted polarized quark distributions.

Related to the Bjorken sum rule, at $Q^2=0$ there is another sum rule for 
spin structure, the Gerasimov-Drell-Hearn (GDH) sum rule~\cite{gdh}. 
A generalized GDH sum rule~\cite{ggdh} 
connects the GDH sum rule with the Bjorken sum 
rule and provides a clean way to test theories with experiments over the
entire $Q^2$ range. 
JLab experiments measured the 
generalized GDH sum in the low to intermediate
 $Q^2$ region. These results and the results on the moments of the
spin structure functions provide a bridge between the quark-gluon
picture at high $Q^2$ to the coherent hadronic picture at low $Q^2$.
In particular, at the low end of the $Q^2$ range, the 
results were compared with Chiral Perturbation Theory calculations and provided
tests of the current understanding of the chiral dynamics of QCD.

\section{Inclusive Polarized Electron-Nucleon Scattering}

For inclusive polarized electron scattering off a polarized 
nucleon target, 
the cross section depends on four structure functions, $F_1(Q^2,x)$, 
$F_2(Q^2,x)$, $g_1(Q^2,x)$ and $g_2(Q^2,x)$, where 
$F_1$ and $F_2$ are the unpolarized structure functions 
and $g_1$ and g$_2$ the polarized structure functions. 
In the quark-parton model, 
$F_1$ or $F_2$ gives the quark momentum 
distribution and $g_1$ gives the quark spin distribution.
Another physics quantity of interest is the virtual photon-nucleon 
asymmetry $A_1$
\begin{equation}
A_1={g_1-(Q^2/\nu^2) g_2 \over F_1} \approx {g_1 \over F_1}.
\end{equation}

\section{Spin Asymmetries in the High-$x$ Region}

To a first approximation, the constituent quarks in the nucleon are
described by the SU(6) wavefunctions.
The SU(6) symmetry leads to the following predictions: 

\begin{equation}
A_1^p=5/9;\ \ A_1^n=0; \ \ \Delta u/u=2/3; \ \ 
\Delta d/d=-1/3; \ \ \ {\rm and} \ \ \ F_2^n/F_2^p=2/3.
\label{eq:SU6}
\end{equation}

These predictions are not expected to work
in the low-$x$ region because the sea quarks and gluon contributions are not
included.
Experimental data on $F_2^n/F_2^p$ agree poorly with the SU(6) quark model 
predictions even in the high-$x$ region ($ x >0.4$), which is a sign that 
SU(6) symmetry is broken for valence quarks. However, a recent 
analysis~\cite{dumodel} revealed a possible large uncertainty 
associated with the nuclear corrections to the $F_2^n/F_2^p$ data in 
the high-$x$ region, which makes the observed SU(6) breaking in $F_2^n/F_2^p$ 
less significant.
 
Relativistic Constituent Quark Models (RCQM) with broken SU(6) symmetry, e.g., 
the hyperfine 
interaction model~\cite{vqm}, lead to a dominance of a `diquark' 
configuration~\cite{diquark} 
with the diquark spin $S=0$ at high $x$. This implies that as $x\rightarrow1$:
\begin{equation}
 A_1^p\rightarrow 1;\ \
   A_1^n\rightarrow 1;\ \ \Delta u/u \rightarrow 1;\ \ {\rm and} \ \ 
\Delta d/d \rightarrow -1/3.
\label{eq:rnpqcd}
\end{equation}
\noindent In the RCQM, the relativistic effect takes into
account the quark orbital angular momentum and reduces the valence quark 
contributions to the nucleon spin from 1 to about $0.7 - 0.75$.
 
Another approach is with leading-order pQCD~\cite{pQCD}, which assumes the 
quark orbital angular momentum is negligible and leads to hadron helicity 
conservation. 
It yields the same limiting values for $A_1^p$ and $A_1^n$ as previously, 
but different limiting values for $\Delta u/u$ and 
$\Delta d/d$:  

\begin{equation}
\Delta u/u \rightarrow 1;\ \ {\rm and} \ \ 
\Delta d/d \rightarrow 1.
\label{eq:rnppqcd}
\end{equation}
\noindent
Not only are the limiting values at $x\rightarrow 1$ important, but also
the behavior in the high-$x$ region. How $A_1^n$ and  $A_1^p$ 
approach their limiting values when $x$ approaches 1 is sensitive to
the dynamics of the valence quarks. 

In 2001, JLab experiment E99-117~\cite{e99117} was carried out in Hall A
to measure $A_1^n$ with high precision in the $x$ region from 0.33 to 0.61
($Q^2$ from 2.7 to 4.8 GeV$^2$). 
Asymmetries from inclusive scattering of 
a highly polarized 5.7 GeV electron beam 
on a high pressure ($>10$ atm) (both longitudinal and
transversely) polarized $^3$He target were measured. 
Beam polarization was measured with a {M\o}ller polarimeter
 and a Compton polarimeter. The
average beam polarization was $78\% \times (1 \pm 0.03)$. 
The $^3$He target was polarized by spin exchange with optically pumped Rubidium. The average in-beam polarization
was $40\% \times (1 \pm 0.04)$. The scattered electrons
were detected with two high-precision spectrometers with
their standard detector packages (scintillators
for trigger, vertical drift chambers for tracking, gas Cherenkov counters
and shower counters for particle identification).

Parallel and perpendicular asymmetries
were extracted for $^3$He. After taking into account the beam and target 
polarization and the dilution factor,
they were combined to form $A_1^{^3He}$. Using the most recent 
model~\cite{model}, nuclear
corrections were applied to extract $A_1^n$. Final results on $A_1^n$
are shown in the left panel of Fig. 1. For clarity, not all theoretical 
predictions are shown. A more complete list is given in Ref.~\cite{e99117}

The experiment greatly improved the precision
of data in the high-$x$ region. This is the first evidence that 
$A_1^n$ becomes positive at large $x$, showing clear SU(6) symmetry 
breaking. The results are in good agreement with the LSS 2001 pQCD
fit to previous world data~\cite{LSS2001} (solid curve) and 
the statistical model~\cite{stat} (long-dashed curve).
The trend of the data is consistent with the RCQM predictions
(the shaded band). The data disagree with the predictions from the 
leading-order pQCD models (short-dashed and dash-dotted curves).

Assuming the strange sea quark contributions are negligible in the region
$x > 0.3$, the 
polarized quark distribution functions $\Delta u/u$ and $\Delta d/d$ were 
extracted from our neutron data combined with the world proton data. 
The results are shown
in the right panel of Fig. 1, along with predictions from the RCQM (dot-dashed
curves), leading-order pQCD (short-dashed curves), the LSS 2001 fits 
(solid curves) and the statistical model (long-dashed curves).  The results 
agree well with RCQM predictions as well as the LSS 2001 fits and statistical 
models but are in significant disagreement with the 
predictions from
the leading-order pQCD models assuming hadron helicity conservation. This
suggests that effects beyond leading-order pQCD, such as the quark orbital
angular momentum, may play an important role in this kinematic region.

\noindent
\parbox[t]{0.6\textwidth}{\centering\includegraphics[bb=30 -28 482 455, angle=0,width=0.6\textwidth]{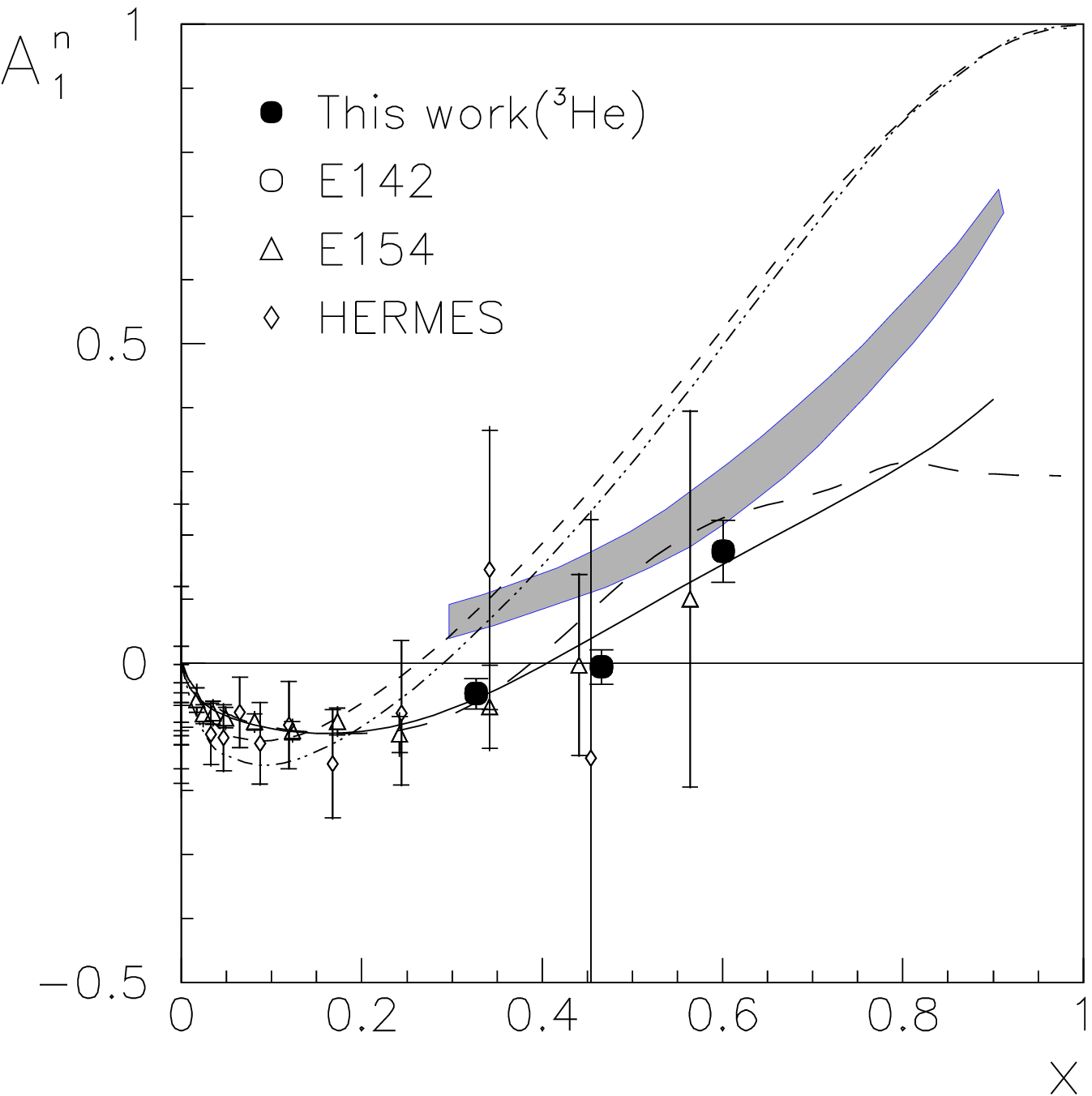}}
\parbox[t]{0.6\textwidth}{\centering\includegraphics[bb=30 -28 482 455, angle=0,width=0.6\textwidth]{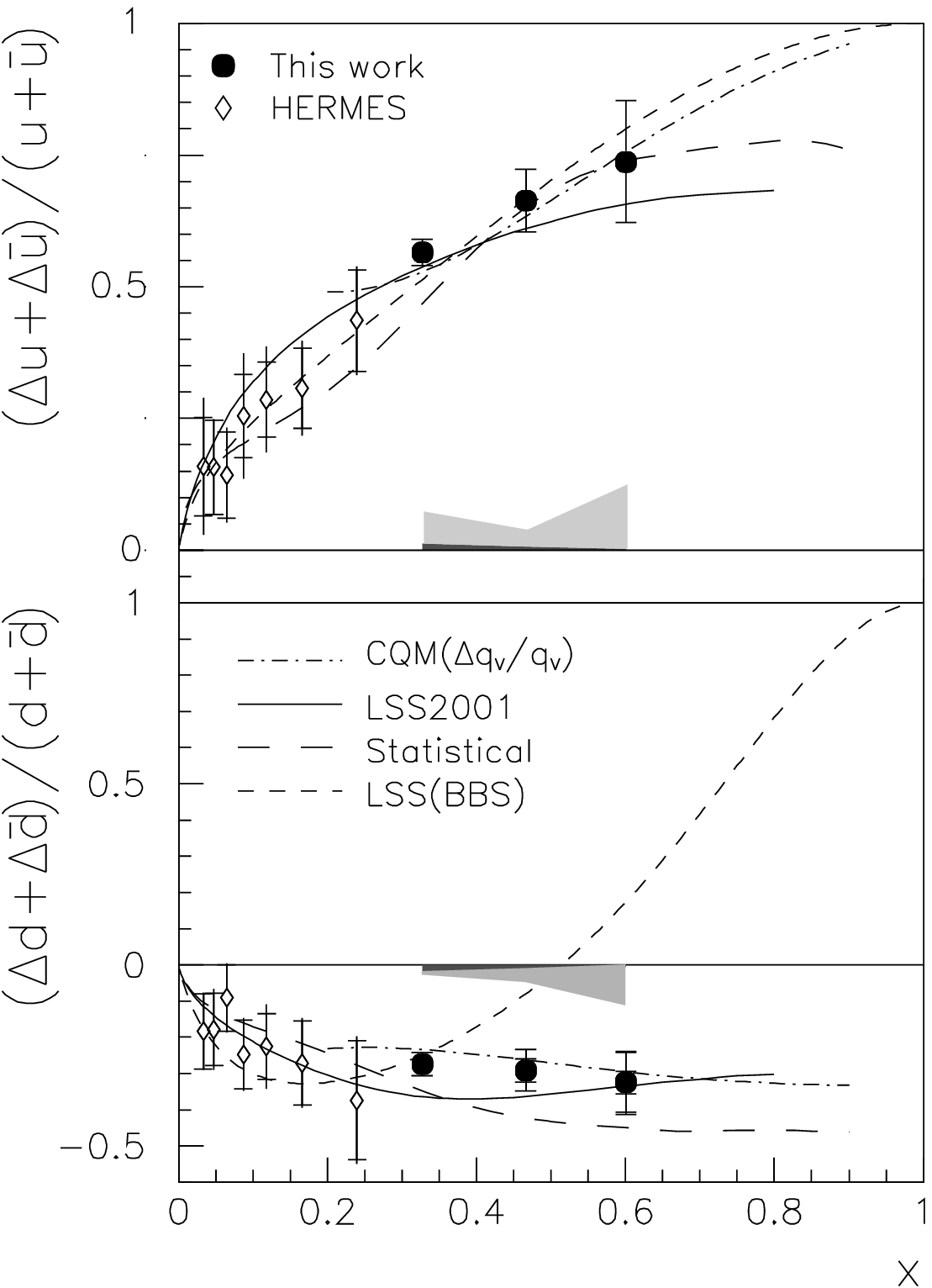}}
 {\centerline{\footnotesize 
        Fig.~1: A$_1^n$, $\Delta u/u$ and $\Delta d/d$ results compared with 
the world data and theoretical predictions.}}

\medskip

\section{Generalized GDH sum and moments of the
spin structure functions}

JLab E94-010~\cite{e94010} measured
the generalized GDH sum and the moments of the neutron spin structure functions
$\Gamma_1$ and $\Gamma_2$
in the low to intermediate $Q^2$ range using a polarized electron beam on 
a polarized $^3$He target. 
The measurement of doubly-polarized inclusive cross sections was performed
at five beam energies from 0.86 to 5.1 GeV 
at a scattering angle of $15.5^\circ$. 
Parallel and perpendicular cross-section differences were obtained,
from which $g_1$, $g_2$ and 
$\sigma_{TT}$, $\sigma_{LT}$ 
for $^3$He were extracted. Interpolation to constant $Q^2$ values was 
performed and
the GDH integrals were formed from pion threshold to $W^2=4$ GeV$^2$. 
Finally, nuclear corrections~\cite{ciofi} were applied, 
to extract the GDH integrals for the neutron.
The results are shown in the left-top panel of Fig. 2. 
 The higher energy contributions (for $W^2$ from
4 to 1000 GeV$^2$) were estimated using the parameterization of Thomas and 
Bianchi~\cite{TB}. 

These data show a smooth but dramatic change in the value of the generalized 
GDH sum
from what was observed at high $Q^2$. While not unexpected from 
phenomenological models, these data illustrate the sensitivity to the transition
from partonic to hadronic behavior. The measured values of the first moment
of 
$g_1^n$ are shown in the left-middle panel of Fig. 2, along with the world data from SLAC and HERMES. Also shown are Chiral Perturbation Theory calculations and several model predictions. 
These data provide a precision data 
base for twist expansion analysis at the higher end of the $Q^2$ range, a check for Chiral Perturbation Theory (ChPT)
calculations~\cite{chpt} at the low end of the $Q^2$ range, and establish an important benchmark against which one can 
compare future calculations (such as Lattice Gauge Theory calculations).
The measured values of the first moment of $g_2^n$ are shown in left-bottom 
panel of Fig. 2.
These results indicate the first validation of the Burkhardt-Cottingham sum 
rule~\cite{BC}, $\Gamma_2=0$.

\medskip

\noindent
\parbox[t]{0.6\textwidth}{\centering\includegraphics[bb=104 122 554 605, angle=-90,width=0.4\textwidth]{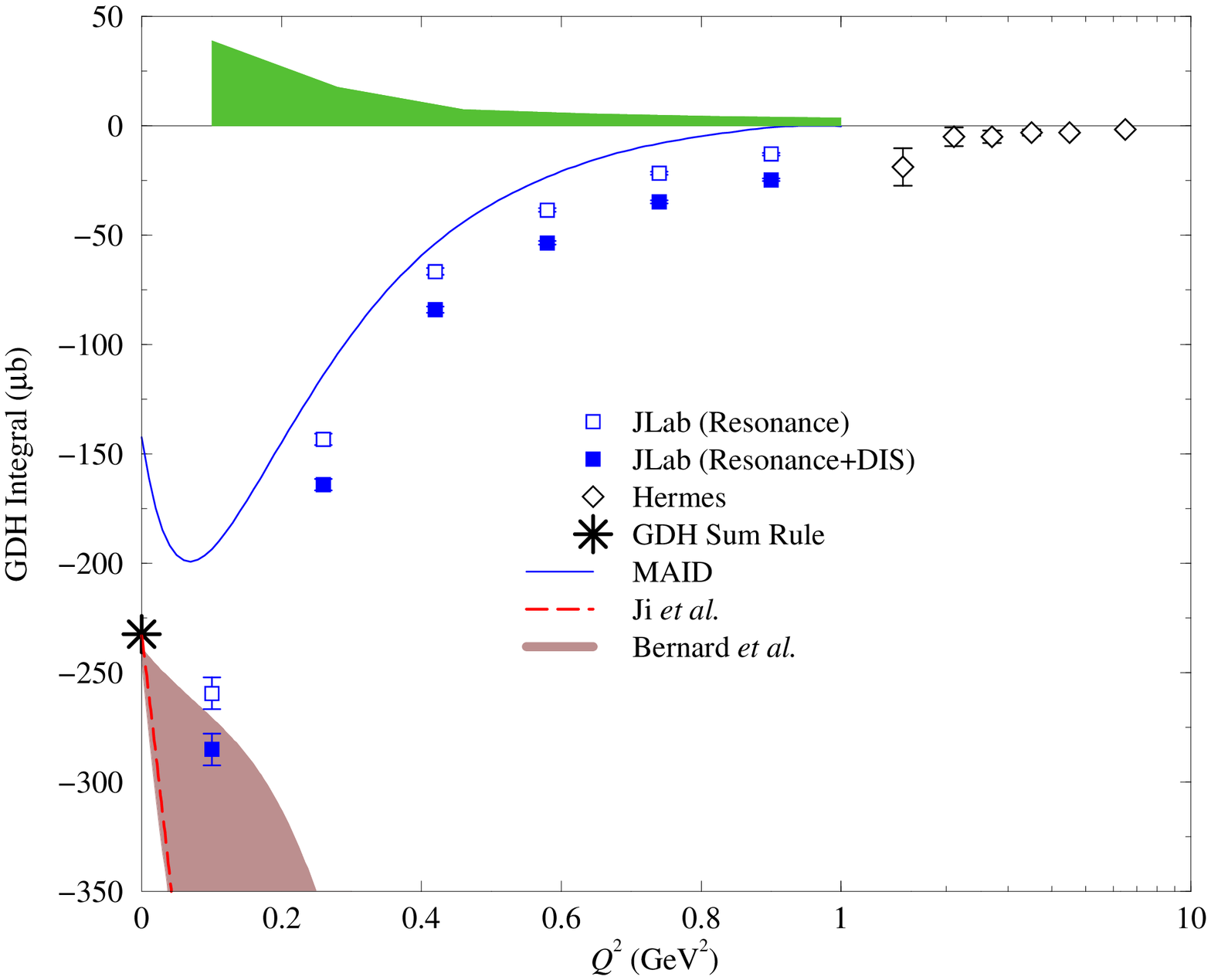}}
\parbox[t]{0.6\textwidth}{\centering\includegraphics[bb=104 222 554 705, angle=-90,width=0.4\textwidth]{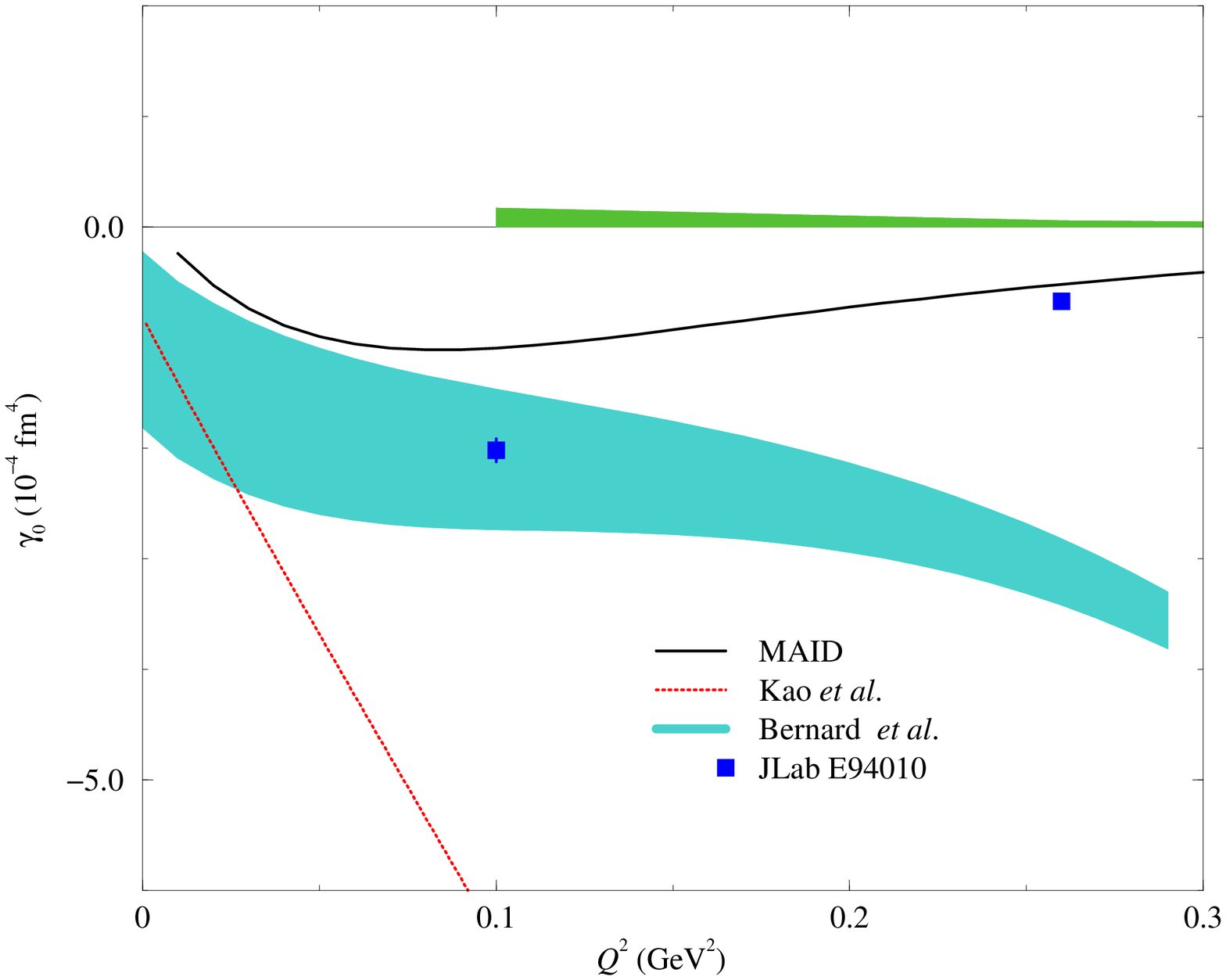}}
\parbox[t]{0.6\textwidth}{\centering\includegraphics[bb=104 122 556 605, angle=-90,width=0.4\textwidth]{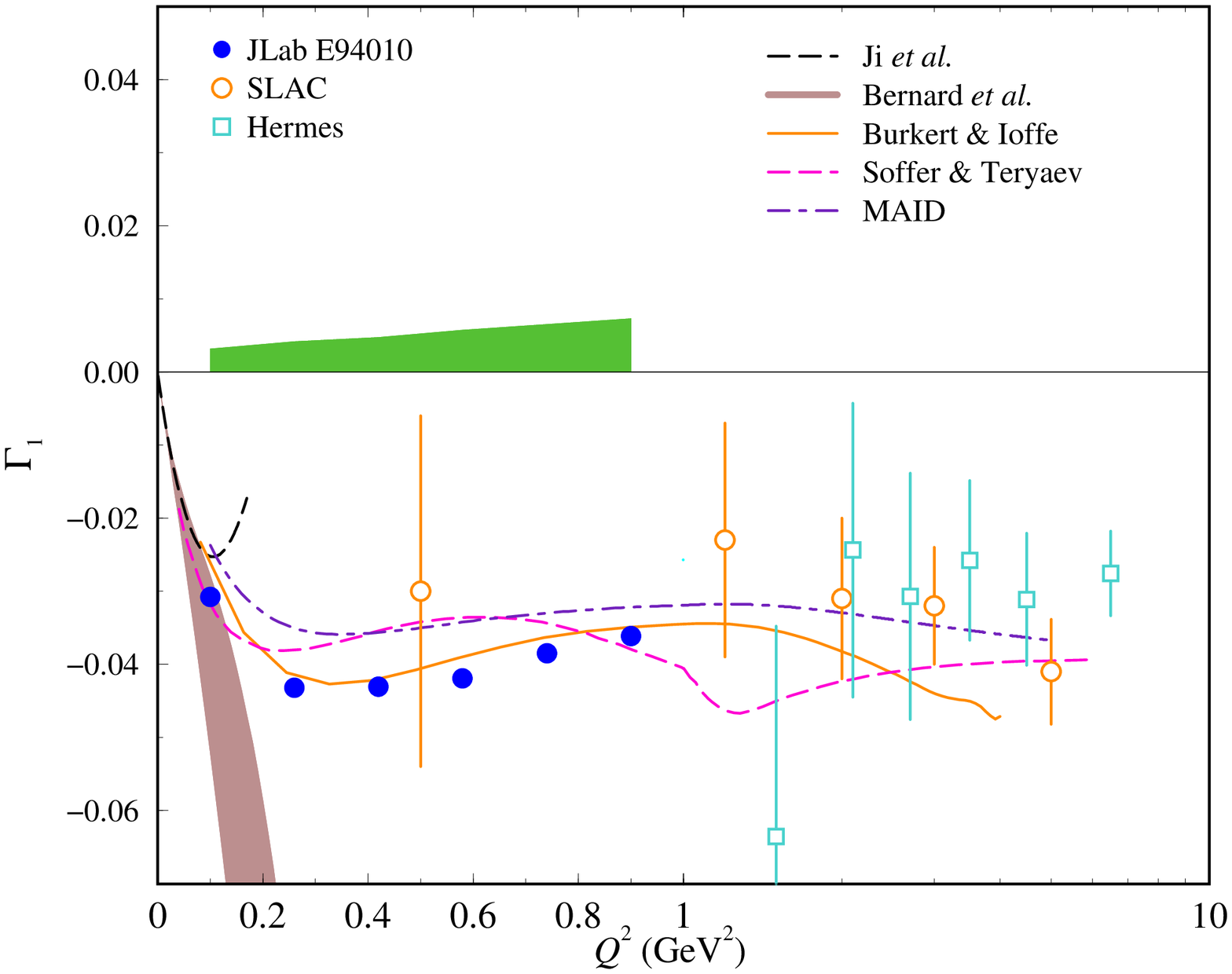}}
\parbox[t]{0.6\textwidth}{\centering\includegraphics[bb=104 222 554 705, angle=-90,width=0.4\textwidth]{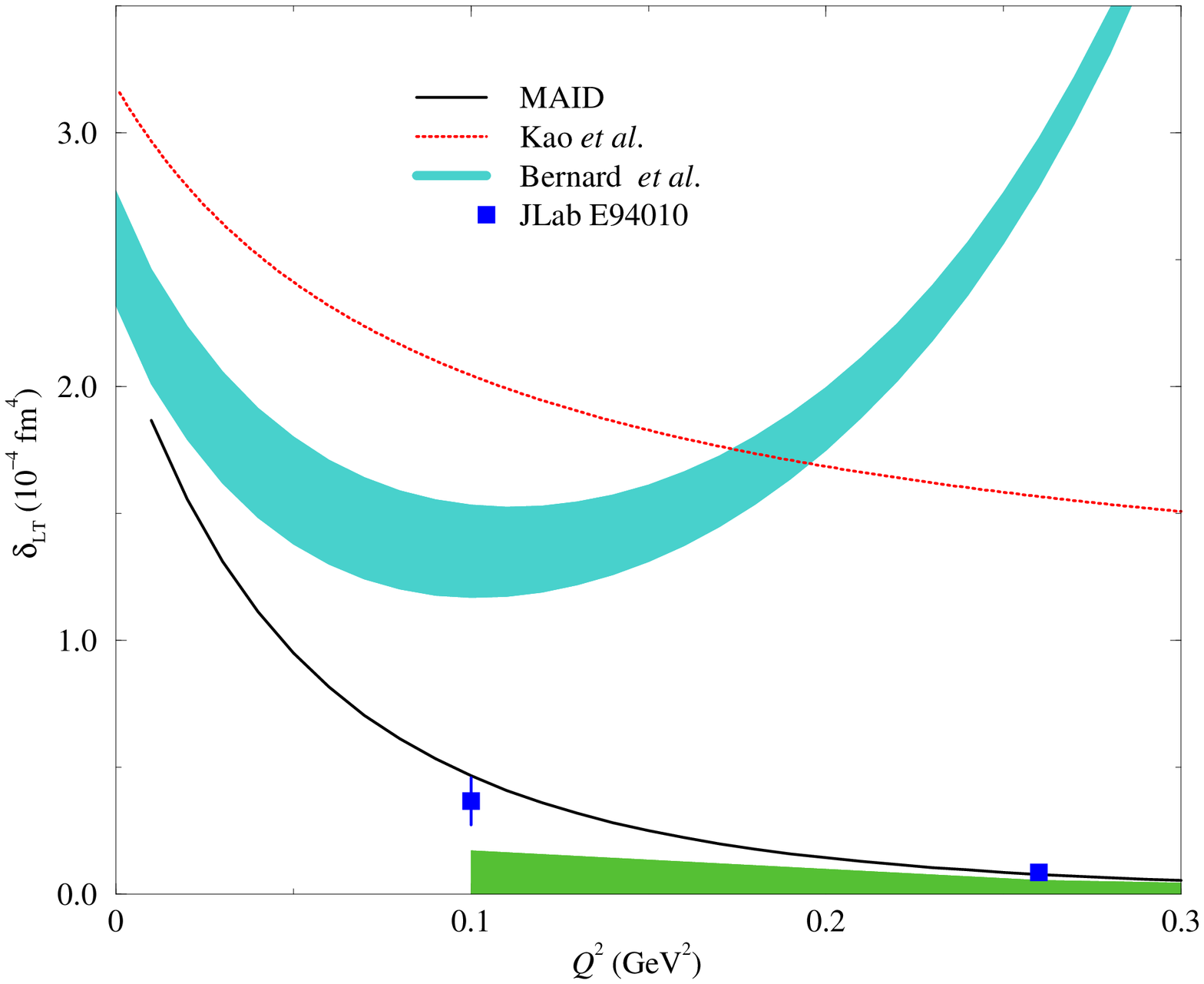}}
\parbox[t]{0.6\textwidth}{\centering\includegraphics[bb=104 122 556 605, angle=-90,width=0.4\textwidth]{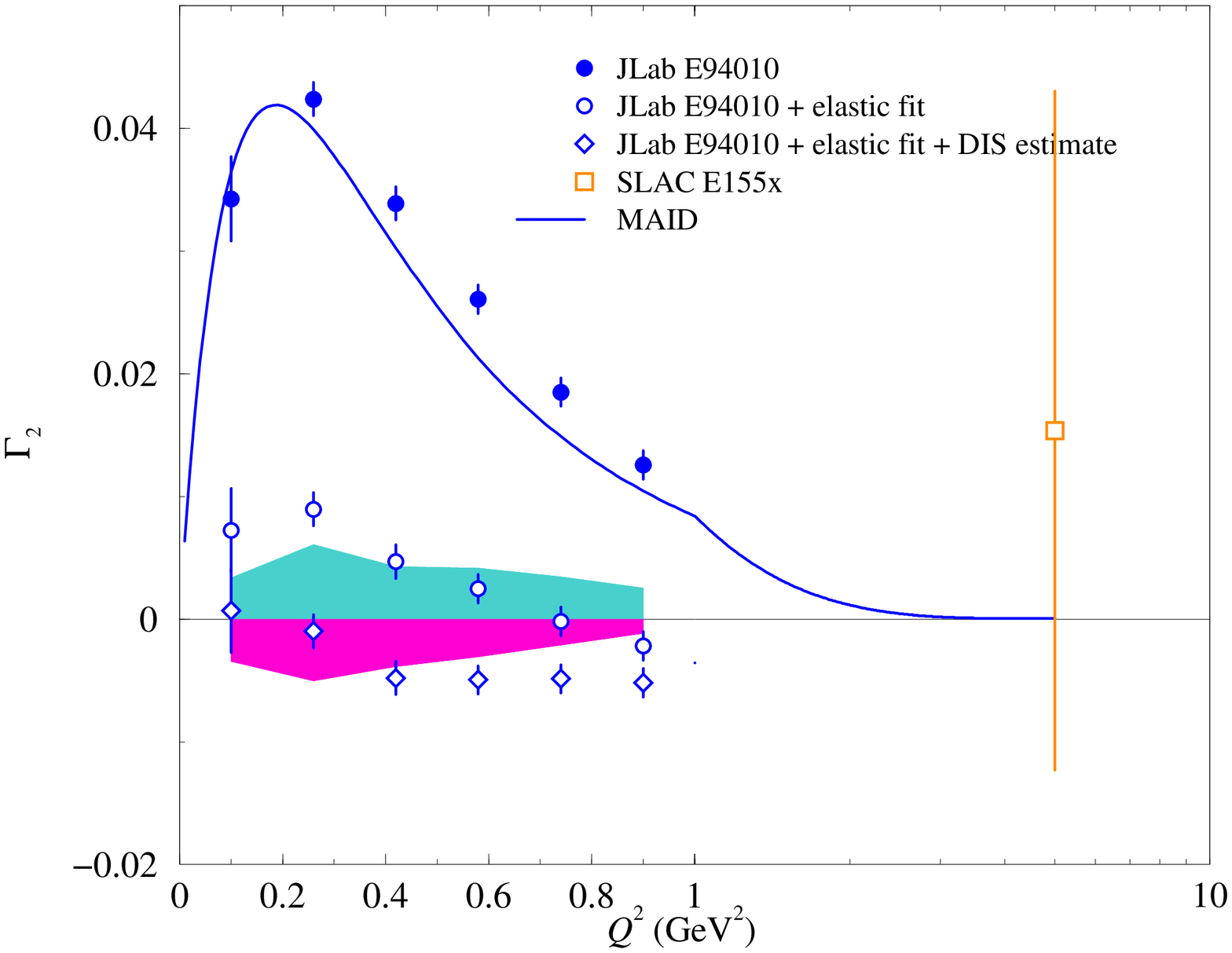}}
\parbox[t]{0.6\textwidth}{\centering\includegraphics[bb=104 222 554 705, angle=-90,width=0.4\textwidth]{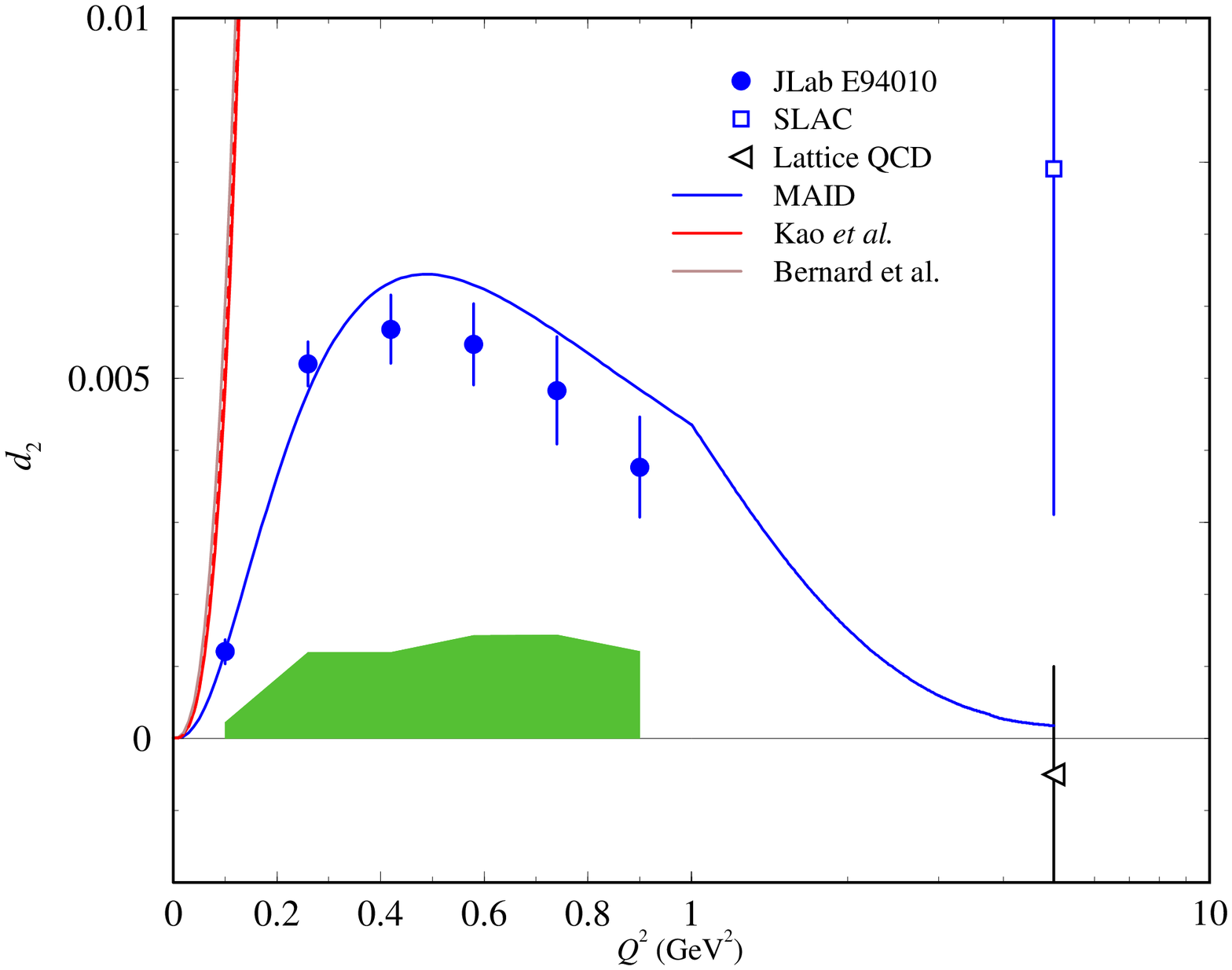}}
{\centerline{\footnotesize 
        Fig.~2: Comparisons of E94-010 results  with world data, 
ChPT calculations and model calculations.}}

\medskip

Higher ($x^2$ weighted) moments of the spin structure functions are related to 
generalized forward spin polarizabilities $\gamma_0$, 
$\delta_{LT}$ 
and the color polarizability $d_2$~\cite{e94010}.
The right panels of Fig. 2 show the E94-010 results on $\gamma_0$, $\delta_{LT}$ and $d_2$, and
the ChPT calculations at low $Q^2$, MAID model~\cite{maid} predictions
at low to intermediate $Q^2$
and the Lattice QCD prediction at high $Q^2$.
The relativistic baryon ChPT with resonance shows good agreement with the 
data for $\gamma_0$ at $Q^2 = 0.1$ GeV$^2$. However, ChPT calculations deviate 
significantly from 
the data for $\delta_{LT}$, which was expected to be an excellent candidate
to check Chiral dynamics of QCD since it was not sensitive to the
dominating resonance ($\Delta$) contributions. This disagreement presents a 
real challenge to theorists.  

A new experiment~\cite{e97110} will extend the generalized GDH
sum measurements to very low $Q^2$ (down to $Q^2=0.02$ GeV$^2$),
below the turn-around point predicted by calculations
(at $Q^2 \approx 0.1$ GeV$^2$). 
ChPT calculations will be extensively tested at low $Q^2$ where they
are expected to be applicable.
Extrapolation to the real photon point
provides an alternative way to test the original GDH sum rule.
Data taking was completed in the summer of 2003. Analysis is underway.
 
In summary, with high polarized luminosity,
JLab has provided a set of high-precision data to study the nucleon
spin structure in a wide kinematic range, which sheds
light on the valence quark structure and helped to
understand the transition region between perturbative and non-perturbative
regions of QCD.  


\begin{theacknowledgments}

The work presented was supported in part 
by the U. S. Department of Energy (DOE)
contract DE-AC05-84ER40150 Modification NO. M175,
under which the
Southeastern Universities Research Association operates the 
Thomas Jefferson National Accelerator Facility.

\end{theacknowledgments}





\end{document}